# Stronger Enforcement of Security Using AOP & Spring AOP

Kotrappa Sirbi, Prakash Jayanth Kulkarni

**Abstract**— An application security has two primary goals: first, it is intended to prevent unauthorised personnel from accessing information at higher classification than their authorisation. Second, it is intended to prevent personnel from declassifying information. Using an object oriented approach to implementing application security results not only with the problem of code scattering and code tangling, but also results in weaker enforcement of security. This weaker enforcement of security could be due to the inherent design of the system or due to a programming error. Aspect Oriented Programming (AOP) complements Object-Oriented Programming (OOP) by providing another way of thinking about program structure. The key unit of modularity in OOP is the class, whereas in AOP the unit of modularity is the aspect. The goal of the paper is to present that Aspect Oriented Programming AspectJ integrated with Spring AOP provides very powerful mechanisms for stronger enforcement of security.Aspect-oriented programming (AOP) allows weaving a security aspect into an application providing additional security functionality or introducing completely new security mechanisms.Implementation of security with AOP is a flexible method to develop separated, extensible and reusable pieces of code called aspects.In this comparative study paper, we argue that Spring AOP provides stronger enforcement of security than AspectJ.We have shown both Spring AOP and AspectJ strive to provide a comprehensive AOP solutions and complements each other.

———————————— ◆ ————————————

## 1 INTRODUCTION

Computer security is a science concerned with the control of risks related to computer use.The accelerating trends of interconnectedness, extensibility and complexity are increasing the threat of such a risk [1]. Application security hardening becomes a priority and one of the fastest growing fields in IT market today. The main goal of application security hardening is to reinforce the application security and therefore to minimize the likelihood of violating security properties.A legitimate question that one could ask is: *"What is the most appropriate computation style or programming paradigm for security hardening?"* A natural answer is to resort to an aspect oriented language. This answer is justified by the fact that aspect oriented languages have been created to deal with the separation of concerns. This is exactly what a security engineer needs when hardening an application. Security engineer would like to inject and strengthen security without digging in the logic of the application/middleware.

*Aspect Oriented Programming (AOP)* [2] has been proposed as a way to improve modularity of software systems by allowing encapsulation of crosscutting concerns. Crosscutting concerns generally refer to non functional properties of software such as *security, synchronization,*

———————————————

- *Kotrappa Sirbi is with K L E's College of Engineering & Technology, Belgaum, 590008, India*

- *Prakash Jayanth Kulkarni is with Walchand College of Engineering, Sangli, India*

*logging, etc.* When programmed, these crosscutting concerns result in tangled and scattered code. In this paper, we address the stronger enforcement of security of AspectJ comes with new concepts and constructs such as: join points, pointcuts, and advices [10].A join point is a point in the control flow graph of an application such as method call, object construction, or field access.A pointcut is a concept that classifies join points in the same way a type classifies values. Accordingly, AspectJ defines various pointcut constructors to designate various join points. An advice is a code fragment executed when join points satisfying its pointcut are reached. This execution can be done before, after, or around a specific join point.

Spring introduces a simpler and more powerful way of writing custom aspects using either a schema-based approach or the @AspectJ annotation style. Both of these styles offer fully typed advice and use of the AspectJ pointcut language, while still using Spring AOP for weaving. the spring supports schema and @AspectJ-based AOP. Spring AOP [3] remains fully backwards compatible with Spring AOP and the lower-level AOP support offered by the Spring APIs of earlier version. AOP is used in the Spring Framework to

- provide declarative enterprise services, especially as a replacement for EJB declarative services. The most important such service is declarative transaction management and
- allow users to implement custom aspects, complementing their use of OOP with AOP.

In this paper, we present a comparative study whether AspectJ alone enough and/or there is usefulness of Spring AOP for stronger enforcement of security in application development.

The remainder of this paper is structured in the following way.The security requirements in enterprise applications is discussed section 2 and section 3 deals with the AspectJ & Spring AOP features for security perspectives, section 4 discuss the implementation of access control and section 5 deals with comparision of AOP AspectJ and Spring AOP security features and their usefulness from a security point of view.

## 2 SECURITY REQUIREMENTS OF ENTERPRISE APPLICATIONS

Security is an important consideration in modern, highly connected software systems.Most applications need to expose functionality through multiple interfaces to allow access to the business data and make complex integration possible. But they need to do so in a secured manner. It isn't a surprise that most enterprises spend substantial time, energy, and money to secure applications. Security is a very important issue in enterprise class information systems, especially in portal/Internet applications. It is related to a possibly huge quantity of sensitive resources in the system and potentially unlimited users access to the application. A violation of security may cause catastrophic consequences for organizations whose security critical information may be disclosed without authorization, altered, lost, or destroyed. There are many standards of security functionality requirements for applications [6], [7], [8], [21]. The most important security functions are:

- **Identification and authentication**: the process of establishing and verifying the claimed identity of user,
- **Access control**: the prevention of unauthorized use of a resource, including the prevention of use of a resource in an unauthorized manner,
- **Accountability**: the property that ensures that the actions of an entity may be traced uniquely to the entity,
- **Audit:** an independent review and examination of system records and activities in order to test for adequacy of system controls, to ensure compliance with established policy and operational procedures, and to recommend any indicated changes in control, policy, and procedures.

In this paper, we focus on authentication and access control functions of security.

## 3 AOP-ASPECTJ AND SPRING AOP SECURITY

### 3.1 Aspect Oriented Programming

*Aspect Oriented Programming* (AOP) is the invention of a programming paradigm developed by the Xerox Palo Alto Research Center (Xerox PARC) in the 20th century [14],[15].It allows developers to separate tasks that should not be entangled with the crosscutting concerns, so as to provide better procedures for the encapsulation and interoperability.The core thought of AOP is to makes a complex system as combinations by a number of concerns to achieve. After demand researched, the concerns are divided into two parts: crosscutting concerns and core business concerns. Core business concern is the needs of the business logic and associated business subsystems, such as financial systems, personnel systems. And crosscutting concern is the needs of the various subsystems business, may be involved in some of the public functions, such as log records, security and so on.

### 3.2 AspectJ and Spring AOP

AspectJ and Spring AOP concepts are:

- **Aspect:** A modularization of a concern that cuts across multiple objects.
- **Join point**: A point during the execution of a program, such as the execution of a method or the handling of an exception. In Spring AOP, a join point always represents a method execution. Join point information is available in advice bodies by declaring org.aspectj.lang.JoinPoint parameter type.
- **Advice:** Action taken by an aspect at a particular join point. Different types of advice include "around," "before" and "after" advice. Many AOP frameworks, including Spring, model an advice as an interceptor, maintaining a chain of interceptors "around" the join point.
- **Pointcut**: A predicate that matches join points. Advice is associated with a pointcut expression and runs at any join point matched by the pointcut (for example, the execution of a method with a certain name).The concept of join points as matched by pointcut expressions is central to AOP.Spring uses the AspectJ pointcut language by default.
- **Introduction**: Also known as an inter-type declaration.Declaring additional methods or fields on behalf of a type. Spring AOP allows introducing new interfaces and a corresponding implementation to any proxied object.
- **Target object**: Object being advised by one or more aspects. Also referred to as the advised object. Since Spring AOP is implemented using runtime proxies, this object will always be a proxied object.
- **Weaving**: Linking aspects with other application types or objects to create an advised object. This can be done at compile time (using the AspectJ compiler, for example), load time, or at runtime. Spring AOP, like other pure Java AOP frameworks, performs weaving at runtime.

Different advice types include:

- **Around advice**: Advice that surrounds a joinpoint such as a method invocation. This is the most powerful kind of advice. Around advices will perform custom behavior before and after the method invocation.They are responsible for choosing whether to proceed to the joinpoint or to shortcut executing by returning their own return value or throwing an exception.

- **Before advice**: Advice that executes before a joinpoint, but which does not have the ability to prevent execution flow proceeding to the joinpoint unless it throws an exception.
- **Throws advice**: Advice to be executed if a method throws an exception.Spring provides strongly typed throws advice, so it is possible to write code that catches the exception (and subclasses) ,without needing to cast from Throwable or Exception.
- **After returning advice**: Advice to be executed after a joinpoint completes normally i.e., if a method returns without throwing an exception.

Around advice is the most general kind of advice. Most interception-based AOP frameworks, such as Nanning Aspects, provide only around advice. As spring, like AspectJ, provides a full range of advice types.The pointcut concept is the key to AOP, distinguishing AOP from older technologies offering interception.Pointcuts enable advice to be targeted independently of the OO hierarchy and also pointcuts provide the structural element of AOP.

AspectJ has a comprehensive and expressive pointcut specification language that allows specifying particular points in the control flow of the program where advices are to be applied. All of them are important from a security standpoint .Table 1 shows the usefulness of the AspectJ pointcuts according to the security target. Although AspectJ supports those efficient and useful pointcut designators for security hardening, they are not enough to express all the security hardening practices. Indeed, the following possible extensions to AspectJ are identified for security hardening:

- Dataflow pointcut.
- Predicted control flow pointcut.
- Loop pointcut.
- Wildcard for pattern matching.
- Modifiers in type pattern syntax.
- Pointcuts for getting and setting local variables.
- Synchronized block join points.

The detailed discussion about the suggested extensions available in [4], [5], [10],[12],[13],[18].

## 3.3 Spring AOP and Security
### 3.3.1 Spring AOP

In Spring AOP aspects are nothing more than regular spring beans, which themselves are plain-old Java objects (POJO) registered suitably with the Spring Inversion of Control container. The core advantage in using Spring AOP is its ability to realize the aspect as a plain Java class. In Spring AOP, a join point exclusively pertains to method execution only, which could be viewed as a limitation of Spring AOP. However, in reality, it is enough to handle most common cases of implementing crosscutting concern. Spring AOP uses the AspectJ pointcut expres-sion syntax [3], [9], [11], [16], [19].

### TABLE 1
### ASPECTJ POINTCUT AND SECURITY

| Security Hardening Target | Pointcut |
|---|---|
| Method call/execution or constructor call/execution. Filed read/write | call, execution get, set |
| Setting security environment during the initialization of a class or an object. | Initialization,staticinitialization,preinitialization |
| Execution context | args,this,target |
| Executing security hardening code depends on a particular condition | if |
| Within a particular class or method In the control flow of other particular points Log exceptions related to security | within,withincode Cflow,pcflow handler |

### 3.3.2 Spring Security

Spring security began in late 2003 as "*The Acegi Security System for Spring*", Acegi Security became an official Spring Portfolio project towards the end of 2007 and was rebranded as "*Spring Security*" [17].Spring Security (formerly known as Acegi) along with AOP can simplify the implementation of the classic crosscutting concern of stronger enforcement of security in an application. Enterprise applications need to address many crosscutting functionalities: transaction management, security, auditing, service-level agreement, monitoring, concurrency control, improving application availability, error handling, and so on. Many enterprise applications use AOP to implement these functionalities. All the examples in given here are based on real-world problems and their AOP solutions.Virtually every project that uses Spring uses AOP [9].Many applications start with prewritten aspects supplied with Spring (primarily transaction management and security).But due to the AspectJ syntax, writing custom aspects is becoming a common task. After reaching the limits of Spring AOP, many applications move toward AspectJ weaving. The typical trigger point for this change is crosscutting of domain objects or other forms of deeper crosscutting functionalities.At that time, it's common to start with the AspectJ syntax (which is used with Spring's proxy-based AOP) along with the load-time weaver. However, applications that don't use spring often use AspectJ weaving from the beginning.

Spring security along with AOP can simplify the implementation of the classic crosscutting concern of securing enterprise application.The security aspect needs to do two things, firstly select join points that need authentication or authorization and then advise the selected join points to perform authentication and authorization as shown in Listing 1.

The interesting piece of code is the computation of the security attributes. In the conventional technique, each method creates a separate security attribute object. In an AOP solution, the same advice applies to all advised join points, yet each join point may require a different attribute.Therefore, the advice may need some cooperation from the advised code or some external configuration to compute a correct attribute at each join point. One way to achieve this collaboration is to use annotations and also security aspects can be applied using either the proxy-based or byte-code based AOP. The security aspect acts as a controller that mediates between the core system and the security subsystem.

```
aspect SecurityAspect {
private AccessDecisionManager accessManager;
pointcut secured () : ...

before() : secured () { //advice
SecurityAttribute sa = ...
accessManager.checkPermission(sa);
}
}
```

Listing 1: Spring AOP Security Aspect

Security requirements vary widely, and many implementations exist to meet these needs [20], [21]. For example, an authentication requirement may vary from simple web-based authentication to a single sign-on solution.Storage for credentials (such as passwords) and authorities (typically roles such as ADMIN or USER) varies widely as well from a simple text file to a database or Lightweight Directory Access Protocol (LDAP).These variations make the already complex topic of security even more so. The increased complexity warrants raising the level of abstraction. But creating such an abstraction is very complex task. This is where Spring Security comes into play [16]. By providing an abstraction layer and an implementation for most commonly used security systems. Furthermore, Spring Security provides ready-made solutions for a few common scenarios that allow implementing certain security requirements by including just a few lines of configuration.

## 4 ACCESS CONTROL TECHNIQUES IN SPRING AOP

Access control is a server–side mechanism.After invokeng a method the decision has to be made whether the method is allowed to be executed or not [20].

*Authentication* is a process that verifies that the user (human or machine) is indeed whom they claim to be. For example, the system may challenge the user with a user-name and password. When the user enters that information, the system verifies it against the stored credentials. Spring Security supports authentication using a wide range of authentication schemes such as basic, forms, database, LDAP, JAAS, and single sign-on. It can also be roll its own authentication to support the specific scheme that the organization is using (and still utilize the rest of the framework, including authorization).After the user credentials have been authenticated, the authenticated user (known as the principal) is stored in the security context.

Figure 1 show the overall structure used in Spring Security for authentication.

*Authorization* is a process that establishes whether an authenticated user has sufficient privileges to access certain resources. For example, only users with the admin privilege may access certain web pages or invoke certain business methods. Spring Security provides role-based and object-level authorization. To accommodate potentially complex custom requirements, it provides several components that can customize. Figure 2 depicts the authorization sequence.

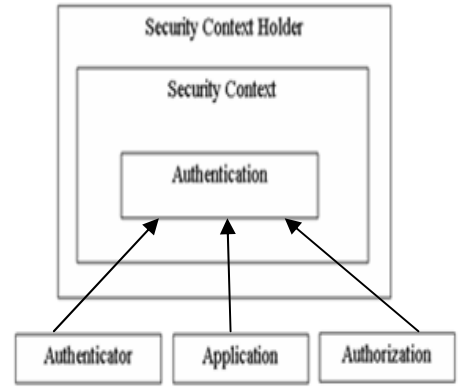

Fig.1. Authentication and Authorization in Spring Security

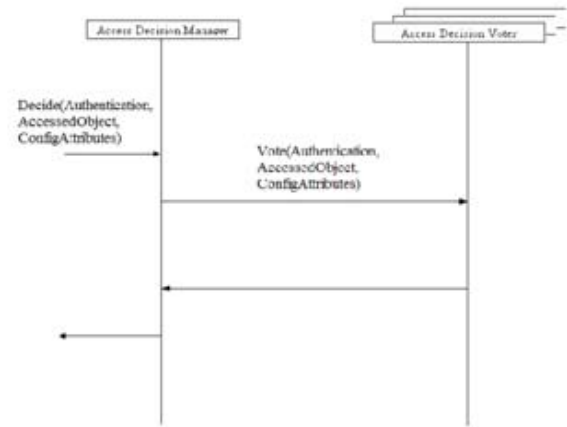

Fig. 2: Spring Security authorization sequence

Implementing security using proxy based Spring AOP shown in Figure 3.Because of similarities due to the common use of AOP as the underlying mechanism. Spring AOP creates a proxy around the service beans, and the security advice ensures authorized access. Spring AOP works only with spring beans.

### 3.1 Spring Security prebuilt solutions

Spring Security provides ready-made solutions that enable developers to secure applications with a few lines of configuration. These solutions target different parts of application: web, service layer, and domain objects.

### 3.1.1 Web security

Securing web applications is a common task, therefore, Spring Security provides special support for this scenario. With namespace-based configuration, a few lines can configure URL level security that ensures that the user has the right authority to access the URLs.

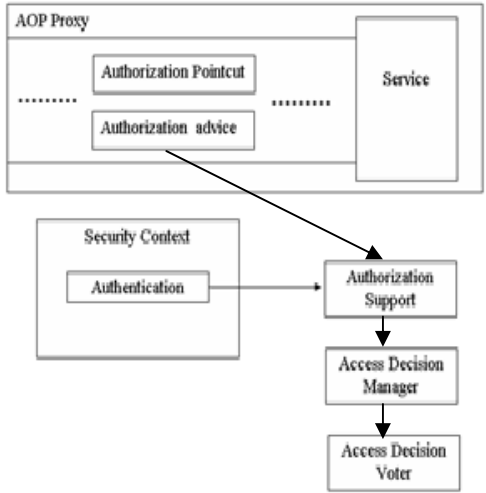

Fig. 3: Securing the service layer using proxy based AOP and Spring Security

For example, the following spring configuration provides authentication using a default login page and ensures that any URLs that end with "delete.htm" are accessible only by users with the ADMIN role. Other URLs are accessed only by users with the USER role.

The declaration of user service to provide username/password Typically, firstly start with a simple snippet as shown in listings 2, modify various attributes, and add additional elements to tailor to the specific needs.

```
<security: http auto-config="true">

<security: intercept-url
pattern="/*delete.htm"
access="ROLE_ADMIN"/> <security:
intercept-url pattern="/**"
access="ROLE_USER"/> </security: http>

Listing 2: Web Security
```

### 3.1.2 Service Level Security

Because most enterprise applications utilize a service layer as the exclusive way of accessing business functionality, it makes sense to secure this layer. Spring Security provides prebuilt aspects along with namespace-based configuration support to secure the service layer with minimal code. It offers two options to specify the access-control information: through the XML-based configuration and through annotations.

## 5    IMPLEMENTATION OF SECURITY

### 5.1 Implementaion of Authentication

To implement security , 3rd party example is used which is party implemented using Acegi. Acegi [17] is a popular and flexible security framework that can be used in enterprise applications.Acegi integrates well with spring and uses spring application contexts for all configurations. Using Acegi Security greatly simplifies implementing authorization in application in a flexible manner. To implement, start with POJO class that will implement the security concern as a Spring AOP[3],[9],[16] aspect and the implementation of method *checkSecurity* (…) (as shown in Listing 3) involves using Acegi APIs and classes such as *SecurityContextHolder* for retrieving the authenticated user and the user's role. Using the user and his role, an application-specific authorization can be implemented.

```
import org.acegisecurity.context.SecurityContextHolder;
import org.acegisecurity.userdetails.UserDetails;
import org.aspectj.lang.ProceedingJoinPoint;
public class MySecurityAspect
{
public Object checkSecurity(ProceedingJoinPoint call) throws
Throwable
{
System.out.println ("from security aspect checking method call for " +
call.toShortString());
Object obj =
SecurityContextHolder.getContext().getAuthentication().getPrincipal()
;
String username = "";
if (obj instanceof UserDetails)
username = ((UserDetails) obj).getUsername (); else
username = obj.toString();
//Do authorization check here
System.out.println ("from security aspect authentic ated user is
"+username);
return call.proceed();
}
}

Listing 3: Spring AOP aspect
```

The important thing to note here is that the entire authorization check lies in a separate aspect (POJO class), which is distinct from the business logic code.This security aspect can be effectively applied to business logic method using the following spring configuration.Firstly, register the regular Java class *SecurityAspect* with spring as a spring bean and then, specify the pointcut and advice:

- The pointcut expression is execution (*com.myorg.springaop.examples.MyService*.*(..) )

- The advice type is "around," and the aspect method name is checkSecurity.

The spring configuration (as shown in Listing 4) for the security aspect,

```
<bean id ="SecurityAspect"
class="com.myorg.springaop.examples.MySecurity
Aspect"/>
<aop:config>
   <aop:aspect ref ="SecurityAspect">
      <aop:pointcut id ="myCutSecurity"
         expression ="execution
(*
com.myorg.springaop.examples.MyService*.*(..))*/
>
      <aop:around pointcut-ref ="myCutSecurity"
method ="checkSecurity"/>
   </aop:aspect>
   ...
   ...
   ...
</aop:config>
```

Listing 4: Spring configuration for the security aspect

Additionally, a spring configuration is needed for configuring Acegi with spring. The configuration uses an in-memory data access object (DAO) provider, in which case, the developer specifies the potential users and roles in the application as simple name-value pairs, as a part of the static spring configuration file. This can easily be inferred from the following spring configuration (as shown in Listing 5):

```
<bean id="authenticationManager" class="org.acegisecurity.providers.ProviderManager">
   <property name="providers">
      <list>
         <ref local="daoAuthenticationProvider"/>
      </list>
   </property>
</bean>
<bean id="daoAuthenticationProvider"
   class="org.acegisecurity.providers.dao.DaoAuthenticationProvider">
   <property name="userDetailsService"><ref bean="inMemoryDaoImpl"/></property>
</bean>
<bean id="inMemoryDaoImpl" class="org.acegisecurity.userdetails.memory.InMemoryDaoImpl">
   <property name="userMap">
      <value>
      kotresh=mypassword,ROLE_TELLER
      sudha=mypassword, disabled, ROLE_TELLER
      </value>
   </property>
</bean>
```
Listing 5: Spring configuration File

## 5.2 Case Study: AspectJ and Spring AOP Integration

Spring's proxy-based AOP framework is well suited for handling many generic middleware and application specific problems. However, there are times when a more powerful AOP solution is required: for example, if programmer needs to add additional fields to a class, or advise fine-grained objects that are not created by the Spring IoC container, in that situation AspectJ is best alternative. Also spring provides a powerful integration with AspectJ.The most important part of the Spring/AspectJ integration allows spring to configure AspectJ aspects using Dependency Injection. This brings similar benefits to aspects as to objects.

- There is no need for aspects to use ad hoc configuration mechanisms; they can be configured in the same, consistent, approach used for the entire application.
- Aspects can depend on application objects. For example, a security aspect can depend on a security manager.
- It's possible to obtain a reference to an aspect through the relevant spring context. This can allow for dynamic reconfiguration of the aspect.

AspectJ aspects can expose JavaBean properties for Setter Injection, and even implement spring lifecycle interfaces such as BeanFactoryAware.In most cases, AspectJ aspects are singletons, with one instance per class loader. This single instance is responsible for advising multiple object instances.A Spring IoC container cannot instantiate an aspect, as aspects don't have callable constructors. But it can obtain a reference to an aspect using the static aspectOf() method that AspectJ defines for all aspects, and it can inject dependencies into that aspect.

Consider a security aspect, which depends on a security manager. This aspects applies to all changes in the value of the balance instance variable in the Account class.It couldn't do in case of using Spring AOP.The AspectJ code for the aspect (one of the Spring/AspectJ samples), is shown below (Listing 6)

```
public aspect BalanceChangeSecurityAspect {
private SecurityManager securityManager;
public void setSecurityManager(SecurityManager
securityManager) {
this.securityManager = securityManager;
}
private pointcut balanceChanged() :
set(int Account.balance);
before(): balanceChanged () {
this.securityManager.checkAuthorizedToModify();
}
}
```

Listing 6:AspectJ Security Aspect(Vs Spring AOP)

This aspect can be configuring in the same way as an ordinary class.

Developer doesn't need to do anything in spring configuration to target this aspect. It contains the pointcut information in AspectJ code that controls where it applies. Thus it can apply even to objects not managed by the Spring IoC container.Using AspectJ pointcuts to target spring advice, we plan to provide the ability for AspectJ pointcut expressions to be used in Spring XML or other bean definition files, to target spring advice. This will allow some of the power of the AspectJ pointcut model to be applied to spring's proxy-based AOP framework. This will work in pure Java, and will not require the AspectJ compiler. Only the subset of AspectJ pointcuts relating to method invocation will be usable.In a latest release of spring will have some spring services packages, such as the declarative transaction management service, as AspectJ aspects. This will enable developer to be used by AspectJ users without dependence on the Spring AOP framework potentially, even without dependence on the Spring IoC container. This feature is probably of more interest to AspectJ users than spring users [20].

## 6 CONCLUSION

Without AOP the functionalities of security would be scattered across the application, with the same code duplicated in different modules. In fact, AOP applied to security solves most of the common practical problems concerning security. We have seen how it is possible to implement these functionalities without having tangled or scattered code, implementing functionalities with aspects and advices.With AOP AspectJ and Spring AOP, configuring it properly for stronger enforcement security of the application thereby we can have cleaner code, much more concise and easier to maintain, debug and adopt. This paper shows how Spring AOP Security up close to reality through implementation of authentication and authorization security crosscutting concerns for strong enforcement of security.

**Kotrappa Sirbi,** M Tech (CSE, 2009), M S (Software System, 1994) and B E (EE, 1984) working with K L E's College of Enginering & Technology Belgaum, India since Dec'1985 and at present working with KLE's BCA, R L S Institute, Belgaum (on deputation).Having two International conferences and two National conferences papers and two journal papers published.Areas of interest are Software Engineering, Object Technology and its evolutions like., Design Patterns, Subject Oriented Programming(SOP) and Aspect Oriented Programming(AOP).Membership of Technical profesional bodies:Life member of ISTE(Indian Society For Technical Education) and member of CSTA (Computer Science Teacher Association),ACM .

**Prakash Jayanth Kulkarni** Ph.D (Electronics, 1993), M E (Electronics, 1986) by Research and B E (Electronics & Telecommunication) working with Walchand College of Engineering, Sangli, India since 1981 and 1980-81 worked with Trans Lines Division, M.S.E.B,having thirteen International conferences papers and nine National Conference papers and six journals paper. His areas of interest are Digital Communication, Digital Image Processing and Computer Vision, Software Engineering, Artificial Neural Network and Genetic Algorithms. In 2001 he received a distinguish *Samaj Shree Award* for rendering services to society.